\documentclass[12pt]{iopart}

\usepackage{psfig}
\begin{document}
\title[Deriving amino acid contact potentials]{Deriving amino acid contact potentials from their frequencies of occurence in proteins: a lattice model study.}
\author{G. Tiana\dag\ddag, M. Colombo\dag, D. Provasi\dag\ddag and R. A. Broglia\dag\ddag * }
\address{\dag Dipartimento di Fisica, Universit\`a di Milano,
Via Celoria 16, I-20133 Milano, Italy.}
\address{\ddag INFN, Sezione di Milano, Via Celoria 16, I-20133 Milano, Italy.}
\address{* The Niels Bohr Institute, University of Copenhagen,
2100 Copenhagen, Denmark.}
\date{\today}

\begin{abstract}
The possibility of deriving the contact potentials between amino acids from their frequencies of occurence in proteins is discussed in evolutionary terms. This approach allows the use of traditional thermodynamics to describe such frequencies and, consequently, to develop a strategy to include in the calculations correlations due to the spatial proximity of the amino acids and to their overall tendency of being conserved in proteins. Making use of a lattice model to describe protein chains and defining a "true" potential, we test these strategies by selecting a database of folding model sequences, deriving the contact potentials from such sequences and comparing them with the "true" potential. Taking into account correlations allows for a markedly better prediction of the interaction potentials.
\end{abstract}

\submitto{\JPC}

\maketitle

\section{Introduction: the "quasi--chemical approximation" revisited} \label{intro}

The idea of obtaining the interaction energy between pairs of amino acids, in the form of a contact potential, from the statistical analysis of known proteins has been developed in detailed by Miyazawa and Jernigan in 1985 \cite{mj}. Assuming that the contact formation can be regarded as a chemical reaction, they make use of Boltzmann statistics to relate the frequencies of occurence of a given pair of amino acids in proteins to the associated contact energy (this is the so--called "quasi--chemical approximation)\footnote{Parallel approaches are those of Mirny and Shakhnovich \protect\cite{mirny}, which optimize the Z--score (a quantity related to the energy gap between the native and the first--excited state of a protein sequence) simultaneously in a large set of different proteins. Vendruscolo {\it et al.} \protect\cite{michele} and van Mourik {\it et al} \protect\cite{amos} perform, with different methods, similar optimization, under a zero--temperature approximation. Of course, several assumption other than the validity of Boltzmann distribution need to be verified, such as the two--body character of the potential \protect\cite{thirum} and its square--well shape \protect\cite{sippl,skolnick,dill}}. This idea has been further developed in refs. \cite{kolinski,mj2,shimada,skolnik2}. 

Since then, this method has undergone wide criticism, both from the point of view of its soundness and of its implementation. Finkelstein and coworkers challenged the validity of the basic assumption of the method \cite{finkelstein}. They highlighted the fact that Boltzmann statistics arises as a consequence of transitions between states of a physical system, while pairs of amino acids are not free to move in the data set of proteins used to obtain the distribution of contacts between pairs of amino acids. In other words, different contacts are not excitations of the same system. Nevertheless, they suggest that contact probabilities still follow a Boltzmann distribution, but for a different reason. Assuming that protein sequences are random, their contacts will be also random, and a Boltzmann--like distribution of pair contacts arises as the ratio between the Gaussian distributions associated with random energies. In this respect, Finkelstein and coworkers interpret the temperature which characterizes this Boltzmann--like distribution as the freezing temperature of the random energy model in conformational space (see also ref. \cite{derrida}). Ben--Naim has questioned whether the energy derived with statistical arguments corresponds to any physical process \cite{bennaim}.
Thomas and Dill have emphasized the fact that, to derive a Boltzmann distribution, one assumes that each pair of amino acids is independent of all other pairs \cite{dill}, an assumption which is not valid because of the high density of amino acids in proteins. Moreover, they noted a possible problem in the fact that frequencies of occurence are obtained counting pairs of amino acids both in the same protein and in different proteins. They also show that the temperature and the reference state (the zero--energy state) are ill--defined (see also refs. \cite{thirum,godzik}).

It is our opinion that the analysis of amino acid pair distributions can be simplified if considered in an evolutionary context. In fact, evolution is the very dynamical process which controls which amino acids are in contact in the native conformations of proteins. The elementary moves of such dynamics are mutations, deletions and insertions, and take place on the timescale of millions of years. The misconception at the basis of the "quasi--chemical approximation" is the idea that the dynamical process which controls which amino acids are in contact is the conformational change which opens or closes a contact, on the timescale of picoseconds.
 This dynamics is indeed present, and controls the stability of a contact (i.e. the probability that it is open rather than closed), but cannot be directly responsible for the likelihood of the amino acids which build the contact. 

In fact, the contact probabilities used to derive the contact energies are those calculated in the native state (i.e. crystallographic or NMR structures) which are, by virtue of Anfinsen's paradigm \cite{anfinsen}, uniquely determined by the protein sequence. Consequently, the energy barrier which a protein must overcome to change a contact with another one is that associated with a change in the sequence (i.e., $\gg$kT, involving the whole replication machinery of the cell), the associated time scale being much longer than that of protein conformational changes (which take place at $kT$). 

From this view point, we can obtain information on the contact frequencies by analyzing the dynamics in the space of sequences at fixed conformation, rather than that in the space of conformations at fixed sequence. If one assumes that evolution has reached an equilibrium state \footnote{This implies that the evolutionary dynamics is ergodic and that it had enough time to converge to an equilibrium state.}, then the contact frequencies observed in proteins can be identified with the equilibrium distribution of contacts associated with the underlying evolutionary dynamics. Whether we can assume that the dataset of known protein sequences represent thermodynamical equilibrium is a difficult issue. A suggestion that this is the case has been made by Rost in ref. \cite{rost}, noting that the distribution of sequence similarities for all known proteins is approximately Gaussian, centered around the random value $1/20$. Note that, however, this suggestion has been questioned in ref. \cite{kolya}.

In order to be quantitative, one has to specify a model for evolutionary dynamics. The simplest model pictures an evolutionary step as a random mutation followed by a check on the ability the new sequence has to fold (i.e., if the mutated protein can still fold to a unique native conformation the move is accepted, otherwise the new sequence is discarded). Moreover, one can add an evolutionary bias toward low--energy proteins, assuming that more stable proteins are selected with higher probability (see below). In other words, the protein dataset is produced by subsequent mutations from a "parent" folding sequence. 

The selection rule, that is checking whether the mutated protein can fold, requires, in principle, a lengthy exploration of conformational space (e.g., by molecular dynamics).
Fortunately, the selection rule can be further simplified. In fact, study of minimal models have shown that the folding ability of a protein sequence is mainly determined by its native state energy \cite{s1,s2,hierarchy}. Approximating the energy distribution of unfolded conformations by mean of the random energy model \cite{derrida} (i.e., assuming that the contact energies in unfolded conformations are independent stochastic variables), it is possible to conclude that if a sequence displays in its conformational ground state an energy lower than a threshold $E_c$, determined by its length and by statistical properties (average and standard deviation) of the interaction potential (thus, independent of the detailed sequence), then it is likely to fold to a unique native state (for a model study, see ref. \cite{hierarchy}). Moreover, the lower is the energy, the more stable is the native state. 

A consequence of these findings is that each point of sequence space can be labeled with a single quantity, namely the conformational ground state energy of the associated sequence, instead that with a whole conformational energy spectrum (that, in principle, would be necessary to assess the folding properties of a sequences). Evolution is then associated with energy fluctuations (i.e., exchange of energy between the protein and an ideal external reservoir), with the constraint that the native energy of a protein can never overcome the threshold $E_c$. In accordance with classical thermodynamics, one can define a temperature $T_{evol}$ which control the equilibrium average native energy and the fluctuations about the average \cite{sh_design}. This temperature, which is not related in any way to the conformational temperature of the protein, has the meaning of an evolutionary pressure toward low--energy proteins.

\begin{figure}
\centerline{\psfig{file=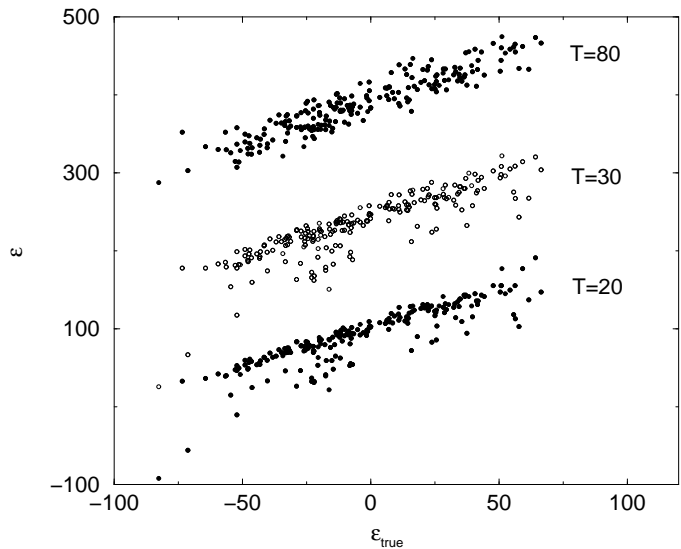,width=9cm,angle=-90}}
\caption{The energies $\epsilon$ predicted from Eq. (\protect\ref{boltzmann}) as a function of the "true" energies $\epsilon^{true}$ used to design the sequences, for three different evolutionary temperatures $T$. Note that the energies are determined except for an additive constant. The correlation coefficients are 0.92, 0.86 and 0.88 for values of $T_{evol}$ of 80, 30 and 20, respectively. The associated slopes are 1.10, 1,10 and 1.02, respectively.}
\label{fig_boltzmann}
\end{figure}

One has now to make three caveats concerning the model of evolution discussed above. First, it takes only into account the folding properties of a protein, but not other important features such as functional requirements, binding sites, etc. Second, only point mutations are allowed implying, among other things, that neither the length of the protein nor the threshold energy $E_c$ change\footnote{A more realistic approach to protein evolution should allow for insertions and deletions, and thus for changes in both the length of the chain and of the energy $E_c$. Such a generalization of the problem can easily be accounted for, although in the following we will consider, for sake of clearness, only chains of fixed length.}. Third, it is important that the concentrations of different kinds of amino acids are constrained, otherwise at low values of $T_{evol}$ the system would display only those amino acids with the lowest interaction energy elements, and the chain would look more like a homopolymer (or oligopolymer) than like a protein. This would not only be unrealistic, since proteins typically display comparable concentration of the different types of amino acids (at least in terms of orders of magnitude), but also would challenge the folding ability of the protein, since the value of $E_c$ depends on the average and standard deviation of the interaction matrix, weighted on the concentration of the amino acids. If such concentrations can vary, the value of $E_c$ cannot be considered sequence--independent and the evolutionary algorithm described above fails. In the worst scenario (i.e., very low $T_{evol}$ with the Miyazawa--Jernigan interaction matrix), the evolutionary algorithm would produce mostly homopolymers composed of cysteine. To avoid this problem, we will fix (in Sect. \ref{css}) the concentration of different kinds of amino acids to their average values (cf. ref. \cite{creighton})\footnote{Another solution is that of labeling each sequence with the Z--score instead that with the native energy, like in ref. \protect\cite{mirny}. However this would make the following calculation more complicated.}.

In what follows we work out the distribution probability of amino acid contacts. Using the standard derivation for closed systems (see, e.g., ref. \cite{greiner}), one can obtain the Boltzmann distribution $p(\{\sigma\})=\exp[-E(\{\sigma\})/T]/Z$ for the probability of occurence of a sequence $\{\sigma\}$, $E(\{\sigma\})$ being the associated native energy (with $E(\{\sigma\}<E_c$). Furthermore, one can consider each contact of the protein as a subsystem which, on the evolutionary timescale, can exchange energy (by mutations) with the rest of the protein, pictured as a heat bath. Making use of the basic approximation that each contact is not correlated to the others, the probability of observing a contact between amino acids of type $\rho$ and $\tau$ at equilibrium is
\begin{equation}
\label{boltzmann}
p(\rho,\tau)=\frac{1}{Z}\exp\left[-\frac{\epsilon(\rho,\tau)}{T}\right],
\end{equation}
where $\epsilon(\rho,\tau)$ is the interaction energy and $T$ is a temperature--like parameter which controls the average energy of the contact. At equilibrium, the temperature of the system is equal throughout, and one can set $T=T^{evol}$.

To be noted that the evolutionary Boltzmann statistics describing the distribution of contacts is different from the "quasi--chemical approximation", because it takes place in the space of sequences and not in that of conformations. It is also different from that derived in ref. \cite{finkelstein}, because it does not assume a protein database populated of random sequences. In particular, the temperature which controls the distribution is not a conformational temperature, but an evolutionary temperature.

\begin{figure}
\centerline{\psfig{file=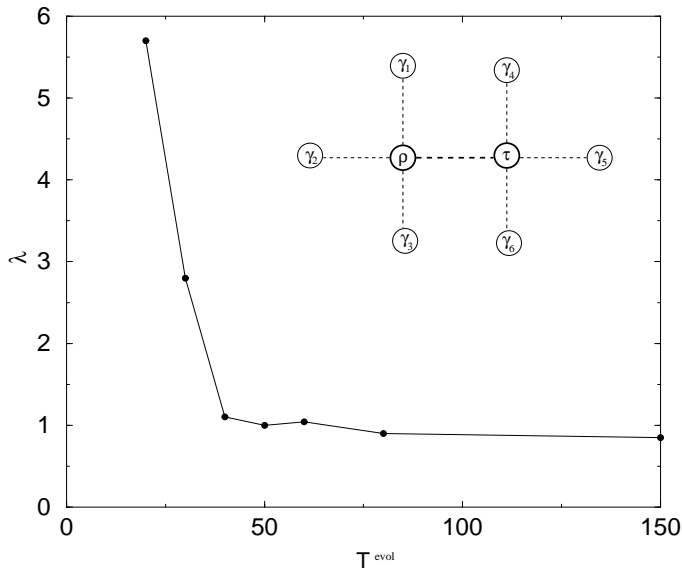,width=9cm}}
\caption{The correlation length at different temperatures for the 80mers model protein interacting through the Gaussian distributed matrix.}
\label{lambda}
\end{figure}

The assumptions made to obtain Eq. (\ref{boltzmann}) are that the system is at equilibrium, that the contacts which build out a protein are uncorrelated and that the concentration of different types of amino acids can vary at will. While the equilibrium condition displays some soundness \cite{rost}, the others are quite crude. In the following, we will discuss corrections to  Eq. (\ref{boltzmann}) to take care of these problems, and test the validity of these corrections within the framework of a lattice model.

\section{The lattice model}

To test the validity of the approximation of Eq. (\ref{boltzmann}) and to develop possible extensions, we follow the approach of refs. \cite{dill,mirny,amos}, where one choses an interaction matrix $\epsilon^{true}(\rho,\tau)$ for model proteins, create a dataset of folding sequences according to this interaction matrix, derive from the sequence dataset the pair frequencies, and from these try to obtain the interaction energies by mean of relationships of the type given in Eq. (\ref{boltzmann}) to be compared with the "true" energies.

The model used for this purpose describes the protein as a chain of beads on the vertexes of a cubic lattice \cite{s1,s2,hierarchy}. We test three kinds of interaction matrices  $\epsilon^{true}(\rho,\tau)$: two of them are extracted from ref. \cite{mj} (Table 6, which we shall call MJ, and Table 5, which we shall call MJ2), and the other one is a symmetric matrix generated extracting its elements at random from a Gaussian distribution. All of them have been shifted and rescaled to display zero mean and standard deviation $0.3$ (in units of $kT^{room}=0.6\;kcal/mol$). A particular feature of the matrix MJ2 is that it displays correlations between its elements, in such a way that can be well approximated by the two lowest eigenvalue terms in a eigenvalue decomposition \cite{wingreen} (naively speaking, most negative terms are concentrated toward one corner of the matrix and the most positive toward the opposite corner). The matrix MJ has been obtained by the matrix MJ2 subtracting the effects of the solvent and does not display any feature which makes it distinguishable from a random matrix. The results obtained with the MJ matrix are substantially identical to those obtained with the random matrix.  

The dataset of sequences has been created fixing one (or more) compact structures and generating sequences with a Monte Carlo algorithm at temperature $T^{evol}$. The pair frequencies $f(\rho,\tau)$ are then calculated counting the relative occurence of the contact between amino acids of kind $\rho$ and $\tau$ in the whole dataset.

\section{Spatial correlations}

Amino acids build complicated networks of interactions within globular proteins, networks which inevitably cause correlations between contacts. As a consequence, Eq. (\ref{boltzmann}) provides a poor estimate of the interaction energies. In Fig. \ref{fig_boltzmann} it is shown the correlation between the "true" energies $\epsilon^{true}$ used to design the sequences (in the case of the Gaussian matrix) and those obtained from Eq. (\ref{boltzmann}), for different choices of the evolutionary temperature. An important parameter to quantify the validity of the energy prediction methods is the root mean square deviation (RMSD, i.e. the root of the reduced $\chi^2$) associated with a linear fit of the predicted energies as a function of the "true" energies, which accounts for the typical variation of the predicted energies from the true ones. This parameter has to be compared with the energy scale of the system, which is $T^{evol}$. The values of RMSD associated with Fig. \ref{fig_boltzmann} are listed in Table 1. 

In order to quantify spatial correlations we define the matrix
\begin{equation}
G_{ij}=\langle {\overline\epsilon_i}{\overline\epsilon_j}\rangle - \langle{\overline\epsilon_i}\rangle\langle{\overline\epsilon_j}\rangle,
\end{equation}
where ${\overline\epsilon_i}$ is the total interaction energy of residue $i$ with its neighbors (i.e., ${\overline\epsilon_i}=\sum_j \epsilon(\sigma_i,\sigma_j)\Delta(|r_i-r_j|)$), while the angular parenthesis indicate the thermodynamic average performed at temperature $T^{evol}$. The elements of the correlation matrix decrease exponentially with respect to the distance $\Delta r\equiv |r_i-r_j|$ of the associated residues, allowing to define a correlation length $\lambda$ as $G(\Delta r)\sim \exp(\Delta r/\lambda)$. The values of $\lambda$ as a function of the evolutionary temperature $T$ are displayed in Fig. \ref{lambda}. 

It has been shown in ref. \cite{hierarchy} that $T^{evol}\approx 30$ is the evolutionary temperature below which good folder sequences are selected.
One can notice that the correlation length $\lambda$ is not negligible below $T^{evol}\approx 30$. This is not completely unexpected, since it is reasonable that good folders, unlike random sequences, display stabilizing long--range effects.

Still lowering the evolutionary temperature below $T\approx 10$, the system enters a phase in which it is difficult to produce equilibrium distributions of sequences with the Monte Carlo algorithm, and which resembles the glassy phase of infinite disordered systems. Since the dynamics in glassy phases is very slow (see, e.g., ref. \cite{bryngelson}), it is unlikely that evolution can have taken place below $T\approx 10$, and consequently we will disregard this phase in the following study. In the following we will develop a simple method to correct Eq. (\ref{boltzmann}) in order to account for such correlations.

If instead of a single contact, one considers a subsystem composed of a contact and all its nearest neighbors (see inset of Fig. \ref{lambda}), the same argument used for the isolated contact suggests that the set of these subsystems follow a Boltzmann distribution, the probability of a contact between amino acids of kind $\rho$ and $\tau$ resulting (see also the Appendix A)
\begin{equation}
\label{p1}
p(\rho,\tau)=\frac{1}{Z}\sum_{\gamma_1...\gamma_6}e^{-\beta[\epsilon(\rho,\tau)+\epsilon(\rho,\gamma_1)+\epsilon(\rho,\gamma_2)+\epsilon(\rho,\gamma_3)+\epsilon(\tau,\gamma_4)+\epsilon(\tau,\gamma_5)+\epsilon(\tau,\gamma_6)]},
\end{equation}
where the sum is over all the neighbors of the two residues of interest and $\beta\equiv 1/T$, where $T$ is the energy scale of the system (which must be provided as an input). Note that Eq. (\ref{p1}) refers, as an example, to a case where each of the amino acids of the pair has 3 further neighbors ($\gamma_1$, $\gamma_2$ and $\gamma_3$ are the neighbors of $\rho$, $\gamma_4$, $\gamma_5$ and $\gamma_6$ are the neighbors of $\tau$), but any number of neighbors is acceptable. Under the approximation that each residue is in contact with the same number $\Gamma$ of other residues, one can rewrite Eq. (\ref{p1}) in the form
\begin{eqnarray}
\label{p2}
\beta\epsilon(\rho,\tau)&=&-\log p(\rho,\tau)+(\Gamma-1)\log\left(\sum_\gamma e^{-\beta\epsilon(\rho,\gamma)}\right)+\nonumber\\
  &+&(\Gamma-1)\log\left(\sum_\gamma e^{-\beta\epsilon(\tau,\gamma)}\right)-\log Z.
\end{eqnarray}
This is a set of $20\times 20$ implicit coupled equations which can be solved numerically to find the interaction energies $\epsilon(\rho,\tau)$. The solution is determined except for the additive constant $\log Z$, which cannot be determined by the set of $p(\rho,\tau)$ alone (see also Sect. VI). The physical meaning of Eq. (\ref{p2}) is to keep into account spatial correlations in the easiest way, assuming a uniform distribution of amino acids in the neighboring sites (see Appendix A) and without the need of calculating three-- or more--body probabilities. Note that the approximation done in Eq.  (\ref{p2}) is not on the range of the correlations, but on the kind of coupling between residues. In fact, the energies which compare at the left side of Eq. (\ref{p2}) are the "true" energies of the system, so that the account for correlations propagate to the whole size of the system.

We first analyze sets of 10000 sequences interacting through the Gaussian matrix designed on a compact conformation built out of 80 residues (that used in ref. \cite{s1}). The dependence of the results on the conformation (or conformations) used will be discussed in Sect. \ref{datab}.

\begin{figure}
\centerline{\psfig{file=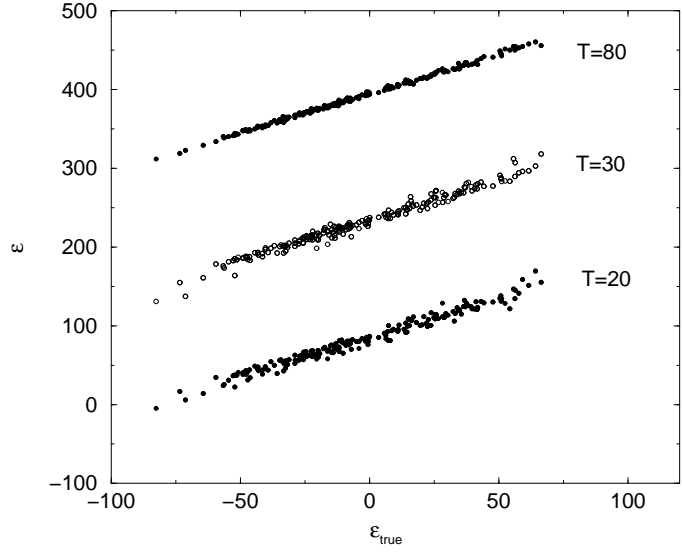,width=9cm,angle=-90}}
\caption{The energies $\epsilon$ predicted keeping into account the spatial correlations through Eq. (\protect\ref{p2}) as a function of the "true" energies $\epsilon^{true}$ used to design the sequences, for three different evolutionary temperatures $T$. The correlation coefficients are 0.987, 0.988 and 0.998, for values of $T_{evol}$ of 80, 30 and 20 respectively. The slopes of the linear fit are 0.99, 1.04 and 0.99, respectively.}
\label{fig_loop}
\end{figure}

\begin{figure}
\centerline{\psfig{file=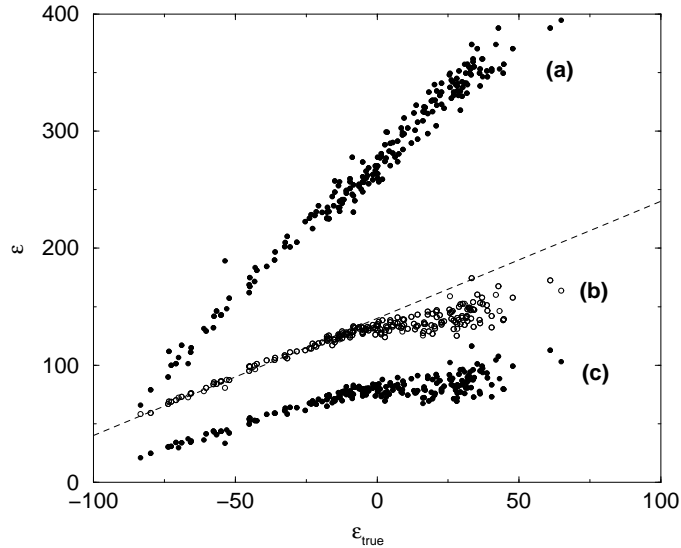,width=9cm,angle=-90}}
\caption{In the case of sequences interacting through the MJ2 matrix selected at $T=30$, it is displayed the energy $\epsilon$ as a function of $\epsilon^{true}$, calculated (a) through Boltzmann statistics (Eq. (\protect\ref{boltzmann})), (b) through the approximation of Eq. (\protect\ref{p2}) and (c) through the approximation of Eq. (\protect\ref{p3}). The dashed line indicates the steepness of the line $\epsilon=\epsilon^{true}$. The correlation coefficients, calculated over the whole set, are 0.98, 0.93 and 0.88, respectively. Restricting the correlation coefficient only to negative energies, it is 0.987, 0.992 and 0.984, respectively.}
\label{fig_mj2}
\end{figure}

The energies calculated solving numerically Eq. (\ref{p2}) through 50000 iterations of a steepest descent optimization algorithm are displayed in Fig. \ref{fig_loop}, as a function of the true energies $\epsilon^{true}$.The associated RMSD is listed in the third column of Table 1. These values have to be compared with the energy scale of the system, that is $T^{evol}$. From these data it is clear that Eq. (\ref{p2}) performs much better than the use of pure Boltzmann statistics. 

A further refinement of the model consists in the keeping also into account of the interaction between nearest neighbors (i.e., between residues $\gamma_1$ and $\gamma_4$ and between $\gamma_3$ and $\gamma_6$ in the inset of Fig. \ref{lambda}, and consequently in Eq. (\ref{p1}) ). The result is
\begin{eqnarray}
\label{p3}
\beta\epsilon(\rho,\tau)&=&-\log p(\rho,\tau)+\log\left(\sum_\gamma e^{-\beta\epsilon(\rho,\gamma)}\right)+\log\left(\sum_\gamma e^{-\beta\epsilon(\tau,\gamma)}\right)-\nonumber\\
&+&\log\left(\sum_{\pi\nu}e^{-\beta(\epsilon(\rho,\pi)+\epsilon(\pi,\nu)+\epsilon(\nu,\tau))}\right)-\log Z.
\end{eqnarray}
This means considering that the amino acids neighboring to the contact of interest are not distributed uniformly, but, more realistically, display a distribution which depend on their own neighbors. The results, also listed in Table 1, show that Eq. (\ref{p3}) is roughly equivalent in the determination of the interaction energies than Eq. (\ref{p2}), although the numerical solution of Eq. (\ref{p3}) is definitely slower.

Similar results are obtained for sequences interacting through the matrix MJ2.
The correspondence between the energies $\epsilon$ calculated in the three different ways discussed above (Eq. (\ref{boltzmann}), Eq. (\ref{p2}) and Eq. (\ref{p3}) ) and the true energies $\epsilon^{true}$ is displayed in Fig. \ref{fig_mj2} for the temperature $T^{evol}=30$. 
One peculiar feature of the results obtained from the MJ2 matrix is that the data associated with the Boltzmann approximation (cf. Fig. \ref{fig_mj2}(a) ), although reasonably correlated, display an effective temperature which is approximately the half of the evolutionary temperature at which the sequences have been selected. This effective temperature could arise from correlations present in the interaction matrix (cf. Appendix B), as suggested by the fact that it does not appear in the case of the random matrix.

\begin{table}
\caption{The RMSD (standard deviation of the residuals) for the case of the Gaussian matrix, calculated in the Boltzmann approximation of Eq. \protect\ref{boltzmann} ("Boltzmann"), in the approximation of Eq. \protect\ref{p2} ("neighbors") and in that of  Eq. \protect\ref{p3} ("loops").}
\begin{tabular}{|c|c|c|c|}
\hline
T & Boltzmann & neighbors & loops\\ \hline
15 & 15.9 &  7.0 & 7.2 \\
20 & 17.7 &  5.0 & 5.3\\
30 & 20.2 &  5.1 & 3.3\\
50 & 15.2 &  1.9 & 2.6\\
80 & 14.0 &  1.7 & 2.6\\
\hline
\end{tabular}
\end{table}

\begin{table}
\caption{The correlations between predicted and true energies for sequences interacting through the MJ2 matrix. The labels are the same as used in Table 1. The value in the column "slope" is the slope of the best fit to a linear function of the data generated by Eq. \protect\ref{boltzmann}. The rows labeled by $(E<0)$ contains the result of fits considering only the negative values of $\epsilon^{true}$.}
\begin{tabular}{|l|c|c|c|}
\hline
 & \multicolumn{2}{|c|}{Boltzmann} & \multicolumn{1}{|c|}{neighbors} \\\hline
T & RMSD & slope & RMSD \\ \hline
15 & 11.9 & 1.7 & 11.9\\
15 $(E<0)$ & 8.8 & &  5.2 \\
20 &  11.4 & 2.3 &  9.2 \\
20 $(E<0)$ &  8.6 & &  4.7 \\
30 & 9.9 & 2.3 & 6.4 \\
30 $(E<0)$ & 8.3 & & 2.2 \\
50 & 10.4 & 2.7 & 5.1 \\
50 $(E<0)$ & 8.4 & & 1.8 \\
80 &  11.5 & 2.9 & 3.8 \\
80 $(E<0)$ &  8.7 & & 1.5 \\
\hline
\end{tabular}
\end{table}

Moreover, for all the three approximations there is a well--defined decrease in the quality of the prediction at $\epsilon^{true}>0$. This is due to a substantial decrease in the statistics associated with the data in this range of energies. The different behavior with respect to the random interaction matrix arises because here some amino acids are intrinsically more attractive than others, due to the order in the interaction matrix. Consequently, at low evolutionary temperatures, these attractive amino acids appear more often than the others. This does not happen in the case of the random matrix, where each amino acid interacts, in average, like any other. In other words, different amino acids display different chemical potentials in the case of the MJ2 matrix, while display the same chemical potential in the case of the random matrix.

The RMSD related to the Boltzmann approximation, listed in Table 2, are calculated using the effective temperature, which corresponds to a rescaling of the actual temperature by a factor indicated in the fourth column of Table 2 (i.e., the slope of the linear fit of the function $\epsilon(\epsilon^{true})$ ). While the features associated with the whole set of data are not very meaningful, being affected by the lack of statistics discussed above, the features associated with the range $\epsilon^{true}<0$ indicate a performance of the prediction methods comparable with that of the random matrix. Note that, for all purposes, the detailed knowledge of the attractive energies is much more critical that the knowledge of the repulsive. 

The values listed in Table 2 for $\epsilon^{true}<0$ indicate that the Boltzmann approximation (Eq. (\ref{boltzmann}) ) is better in the MJ2 case than in the random case. In any case, the approximation of Eq. (\ref{p2}) gives better results in both cases. The behavior of the MJ matrix is identical to that of the random matrix (data not shown).

Note that the accounting for correlations described in Eq. (\ref{p2}) can be easily integrated in any algorithm which derives potentials from pair statistics (e.g., ref. \cite{kolinski,mj2}). Algorithms which optimize simultaneously the parameters which define the potentials, on the other hand, accounts implicitly  for correlations between amino acids in the process of energy optimization. Nonetheless, the explicit treatment of such correlations seems to be fruitful. In the case of the algorithm employed in ref. \cite{mirny}, for example, the correlation coefficient between true and predicted energies is 0.84, while that associated with our method is, in the worst case, 0.88 (see caption to Fig. \ref{fig_mj2}).

\section{Correlations due to amino acid conservation} \label{css}

Looking at the sequences generated at low evolutionary temperatures which produce the data analyzed in the previous Section, one notes an unrealistic feature, namely that at low $T_{evol}$ the frequencies of the different kinds of amino acids are extremely uneven, some of the proteins displaying only a limited subset of all amino acids. Consequently, when working at high evolutionary pressure, could be desirable to constrain the relative concentrations of amino acids.
Of course, in performing this kind of approach, one gives up to obtain the concentration of amino acids as an emergent feature of the system and sets it by hand. The reason why we follow this strategy is that the concentrations of amino acids observed in real proteins are the result of a balance between different factors: not only the thermodynamical requirement of displaying a native energy well below $E_c$, but also the need not to display too many hydrophobic residues to prevent aggregation, to have enough amino acids of a given kind to perform some biological task, etc. The model we employ is too simple to account for all these effects, and accordingly we do not attempt to obtain the concentrations from the model, but we fix it {\it a priori}.

Thus, we have repeated the above calculations using the MJ2 matrix and adding the constrain that the concentration of each amino acid is constant and equal to each other (this is a simplification, since different amino acids occur with different probabilities in real proteins \cite{creighton}. The generalization is straightforward). This constrain introduces nonlocal correlations in the sequence, since the occurence probability of a pair of amino acids now depends on the presence of other amino acids of the same kind everywhere else.

In order to keep into account these constrains, we employ a grand--canonical formalism, introducing chemical potentials $\mu_{\alpha}$ for each kind of amino acid $\alpha$. The Boltzmann (Eq. (\ref{boltzmann})) and loop (Eq. (\ref{p2})) approximations become
\begin{eqnarray}
\label{cons1}
\beta\epsilon(\rho,\tau)&=&-\log p(\rho,\tau)+\mu_{\rho}+\mu_{\tau}-\log Z\\
\label{cons2}
\beta\epsilon(\rho,\tau)&=&-\log p(\rho,\tau)+(\Gamma-1)\log\left(\sum_\gamma e^{-\beta(\epsilon(\rho,\gamma)-\mu_{\gamma})}\right)+\nonumber\\
&&(\Gamma-1)\log\left(\sum_\gamma e^{-\beta(\epsilon(\tau,\gamma)-\mu_{\gamma})}\right)
+\mu_{\rho}+\mu_{\tau}-\log Z,
\end{eqnarray}
respectively. The problem which arise is now the determination of the chemical potentials (we will search for chemical potentials which make each kind of amino acids equiprobable. The generalization is straightforward). For this purpose we will employ a mean field approximation, where 
\begin{equation}
\label{mu}
\mu_{\rho}\approx \frac{1}{20}\sum_{\tau=1}^{20}\epsilon(\rho,\tau),
\end{equation}
that is assuming that inserting an amino acid of given kind into the system costs a free energy given by the average interaction of that amino acid with all other kinds.

\begin{figure}
\centerline{\psfig{file=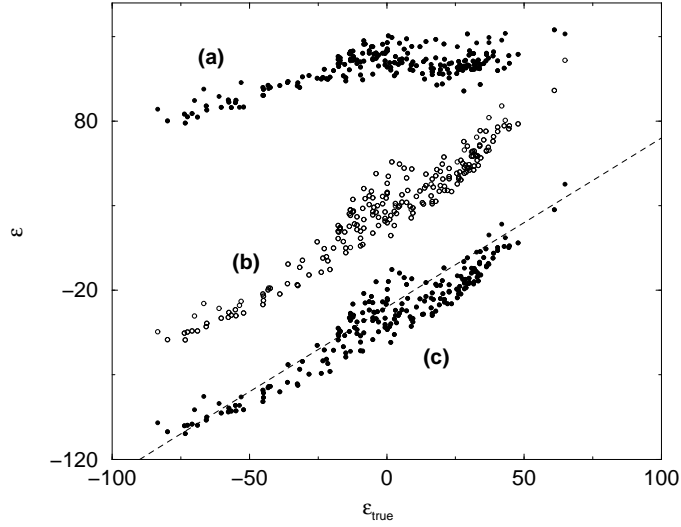,width=9cm,angle=-90}}
\caption{The energies predicted from sequences interacting through the MJ2 matrix and in which the amounts of different kinds of amino acids are kept fixed. The prediction is performed through (a) Eq. (\protect\ref{p1}), (b) (\protect\ref{cons1}) and (c) (\protect\ref{cons2}). The evolutionary temperature is $T_{evol}=30$. The correlation coefficients are 0.66, 0.89 and 0.97, respectively.}
\label{fig_conserved}
\end{figure}

\begin{figure}
\centerline{\psfig{file=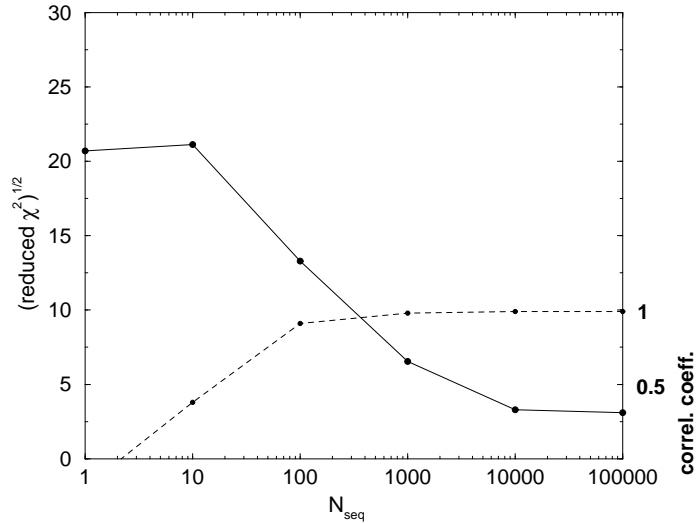,width=9cm,angle=-90}}
\caption{The RMSD (solid curve, left axis) and the correlation coefficient (dashed curve, right axis) as a function of the number of sequences $N_{seq}$ used to collect statistics, using the random matrix and $T=30$.}
\label{fig_stat}
\end{figure}

In Fig. \ref{fig_conserved}(a) it is displayed the result obtained making use of the canonical Boltzmann approximation (Eq. (\ref{boltzmann}) ) for sequences selected at $T^{evol}=30$. The correlation coefficient is very poor (0.66), the slope of the linear fit is 0.23, which gives an effective temperature of $T^{eff}_{evol}=30\cdot 0.23=6.9$. The RMSD is 7.9, that is $1.14$ times the effective temperature. In Figs.  \ref{fig_conserved}(b) and (c) are displayed the results obtained with the gran--canonical formalism of Eqs. (\ref{cons1}) and (\ref{cons2}), respectively, with the chemical potentials calculated from Eq. (\ref{mu}). While the grand--canonical Boltzmann approximation (\ref{cons1}) gives a correlation coefficient of 0.89, the grand--canonical loop approximation of Eq. (\ref{cons2}) gives a correlation coefficient of 0.97. The slopes of the linear fits are 1.02 and 0.98, respectively. The associated RMSD are 7.9 and 9.0, respectively, which are approximately one fourth of the energy scale of the system.

Although the approximation of Eq. (\ref{cons1}) performs much better than the canonical Boltzmann approximation (Eq. (\ref{boltzmann}) ), it performs essentially as well as the approximation of Eq. (\ref{cons2}). This could be due to the fact that the correlations arising from amino acids conservation dump the correlations associated with local proximity of amino acids.

\section{Database dependence of the results} \label{datab}

In order to study what is the amount of sequences needed to obtain a good estimate of the interaction energies, we have calculated the RMSD and the correlation coefficient as a function of the number of sequences $N_{seq}$ used. The energies are calculated by mean of Eq. (\ref{p2}) at $T=30$. The plot, shown in Fig. \ref{fig_stat}, displays a marked worsening of the predicting capability of the algorithm around $N_{seq}\approx 100$.
Since each sequence displays $105$ contacts between its amino acid, it means that one needs $\approx 10^4$ contacts to obtain reliable results. Note that at $N_{seq}\approx 100$ there are $9$ pairs of amino acids out of 210 which never appear in the statistics, a number which grows to 101 for $N_{seq}\approx 10$ and to 162 for $N_{seq}\approx 1$. All these pairs display large, repulsive "true" energies.

When studying real protein sequences one usually does not collect pair statistics from many sequences\footnote{Of course these should be statistically independent sequences, that is, in bioinformatical language, non--homologous sequences.} folding to the same conformation, but rather from sequences folding to different conformations. We have studied a set of 100 sequences, optimized on 10 different 80mer compact conformation (10 sequences optimized on each conformation). The RMSD results 13.1 and the correlation coefficient $0.90$, to be compared with 13.3 found using a single conformation. This suggests that there is no difference in using proteins with different native conformations to collect pair statistics.

Repeating the same calculation using model proteins of different length (27, 36, 48 and 80) provides a RMSD of 16.2 and a correlation coefficient of $0.86$. This slight under-performance appears to be connected with the fact that the value of $\Gamma$ in Eq. (\ref{p2}) depends on the length of the protein, through the ratio of bulk/surface residues, and consequently cannot be considered fixed. This problem becomes more serious in the case of real proteins, where the continuous character of the degrees of freedom makes the coordination number of residues more variable than in the case of lattice models. However, one could expect that this is partially compensated by the fact that real proteins are typically longer than the lattice chains we have employed, and consequently the ratio of bulk/surface residues is larger (at least, for globular proteins). In the case of the set of proteins used by Miyazawa and Jernigan \cite{mj}, the mean coordination number is $10$ and the standard deviation is $4$. To reduce the standard deviation one can, following ref. \cite{mj}, consider only interior residues, that is residues within 7$\AA$ from the centre of the protein, obtaining a mean coordination number of $6$ with a standard deviation of $1.7$ (to be compared with the mean and the average associated with lattice model proteins, that is $4$ and $1.7$, respectively).

To model evolution, we have assumed a constant evolutionary temperature, which also sets the energy scale of the resulting interaction matrix. While that lack of knowledge on the detailed value of this temperature causes no problem in the efficiency of Eq. (\ref{p2}), since the energies are simply rescaled according to $T^{evol}$, problems arise if the set of sequences is selected making use of different temperatures (which is quite a realistic situation). Using a dataset composed of 333 sequences selected at $T^{evol}=20$, 333 sequences selected at $T^{evol}=30$ and 333 sequences selected at $T^{evol}=50$, we obtain an effective temperature of $T^{evol}=24.7$, a RMSD of 7.3 (instead of 3.3, see Table 1) and a correlation coefficient of 0.95 (instead of 0.99).

\section{Conclusions}

We have shown that, looking at the set of proteins from an evolutionary point of view, it is possible to obtain contact energies between amino acids in a sound way, making use of standard thermodynamics (although used in the context of finite systems). This approach also allows to understand which are the limitations in the approximations of the method, as, for example, the presence of correlations between amino acids, and to develop corrections to those approximations.

For sequences generated without any constrain on their composition, accounting for correlations allows to decrease the RMSD from 20.2 to 5.1 in the case of the MJ matrix and from 8.3 to 2.2 in the case of the negative elements of the MJ2 matrix (the correlation coefficients increasing from 0.86 to 0.99 and from 0.98 to 0.99, respectively). Costraining the composition of proteins, the RMSD (relative to the energy scale $T_{evol}^{eff}$) decreases from 1.1 to 0.2 in both the case of the grand--canonical independent--contact approximation and of the grand--canonical formalism with correlation corrections (the correlation coefficients being 0.66, 0.89 and 0.97, respectively).

Of course, there are still strong approximations in this approach, as the step--shape of the potential and its isotropic character. On the other hand, these are computational limitations which do not undermine the soundness of this approach, being easy to extend it to situations displaying more complicated shapes of the potential function (see, e.g., ref. \cite{sippl}). Another problem which has not been dealt with is the determination of the zero of the energies (i.e., the quantity $-T\log Z$). This is a key issue when one wants to use these potentials to determine, for example, binding constants, but it is less critical in folding studies. Anyway, the determination of this parameter is something which cannot be done from the contact probabilities alone, but requires further ingredients, and consequently goes beyond the purpose of the present work.

It is interesting to repeat the calculations performed by Miyazawa and Jernigan, deriving the frequencies of occurence of amino acid pairs in the same database of real proteins used by them \cite{mj}. Comparing the energies obtained making use of the Boltzmann--statistics approach (i.e., Eq. (\ref{boltzmann}), in the spirit of the derivation of ref. \cite{mj}\footnote{Disregarding in this comparison the corrections done in ref. \protect\cite{mj} to account for the effects of the solvent.} with the corrections discussed above, we obtain energies which are substantially different from theirs. Taking care of the spatial correlations (Eq. (\ref{p2}) ), we obtain a set of energies whose correlation parameter with the Miyazawa--Jernigan set is 0.91 and whose RMSD is $0.28\, T$. In keeping into account also the correlations due to amino acids conservation (Eq. (\ref{cons1})) the correlation coefficient becomes 0.96 and the RMSD $0.17\, T$.

\appendix
\section{}

If one assumes that a protein is a closed system, the equilibrium probability for a sequence $\sigma_1,\sigma_2,...,\sigma_N$ is given by
\begin{equation}
\label{a1}
p(\{\sigma_i\})=\frac{1}{Z}\exp[-\beta\sum\epsilon(\sigma_i,\sigma_j)\Delta(|r_i-r_j|)],
\end{equation}
where $\Delta(|r_i-r_j|)$ is a contact function which contains the information about the coordinates $\{r_i\}$ of the (fixed) native conformation. Since we are interested in the probability of a single contact (say, between $\sigma_i$ and $\sigma_j$), we can make the interaction $\epsilon(\sigma_i,\sigma_j)$ explicit, as
\begin{equation}
p(\sigma_i,\sigma_j)=\frac{1}{Z}e^{-\beta\epsilon(\sigma_i,\sigma_j)}\sum_{\sigma_k,\sigma_l,...}\exp[-\beta(\epsilon(\sigma_i,\sigma_k)+\epsilon(\sigma_k,\sigma_l)+...)].
\end{equation}
Focusing our attention on the nearest neighbors (e.g., $k$) of the $i$th and $j$th contacts, we obtain
\begin{eqnarray}
p(\sigma_i,\sigma_j)&=&\frac{1}{Z}e^{-\beta\epsilon(\sigma_i,\sigma_j)}\sum_{\sigma_k}e^{-\beta\epsilon(\sigma_i,\sigma_k)}\nonumber\\
  &&\sum_{\sigma_l,\sigma_m,...}\exp[-\beta(\epsilon(\sigma_k,\sigma_l)+\epsilon(\sigma_l,\sigma_m)+...]+...
\end{eqnarray}
which can be shortened as
\begin{equation}
p(\sigma_i,\sigma_j)=\frac{1}{Z}e^{-\beta\epsilon(\sigma_i,\sigma_j)}\sum_{\sigma_k}e^{-\beta\epsilon(\sigma_i,\sigma_k)}q(\sigma_k)+...
\end{equation}
where $q(\sigma_k)\equiv \sum_{\sigma_l,\sigma_m,...}\exp[-\beta(\epsilon(\sigma_k,\sigma_l)+\epsilon(\sigma_l,\sigma_m)+...)]$ is proportional to the probability that the $k$th site of the protein is occupied by amino acid $\sigma_k$ due to the effect of all other residues of the protein except that in the $i$th site.

The approximation of Eq. (\ref{p3}) consists in assuming that $q(\sigma)=1$, meaning that each kind of amino acid has the same probability of occupying a site in contact with the pair $i-j$ under study.

\section{}

The existence of an effective temperature arises from the the correlations present in the MJ2 correlation matrix. To illustrate this point, consider the example of an interaction matrix which emphasizes the hydrophobic character of the amino acids, in the form  $\epsilon(\rho,\tau)=x(\rho)+ x(\tau)$, then Eq. (\ref{p1}) becomes 
\begin{equation}
p(\rho,\tau)=\frac{1}{Z}\exp\left[ -\frac{\Gamma (x(\rho)+x(\tau))}{T_{evol}}\right]\left(\sum_\gamma e^{-x(\gamma)/T_{evol}}\right)^{2(\Gamma-1)},
\end{equation}
which approximates a Boltzmann distribution with an effective temperature $T^{eff}_{evol}\equiv T_{evol}/\Gamma$ smaller than the real temperature $T_{evol}$. 

This is the simplest kind of correlation which could affect the system. It has been suggested in ref. \cite{wingreen} that the interaction matrix MJ2 has the form  $\epsilon(\rho,\tau)=x(\rho)+ x(\tau)+\lambda x(\rho)\cdot x(\tau)$, corresponding to considering only the first two terms in the Eigenvalue decomposition of the matrix. This, or more complicated, kinds of correlation cannot be accounted for analytically, but are likely to be responsible for the appearance of an effective temperature.

\newpage

\end{document}